\documentclass[12pt]{article}
\usepackage{epsfig}
\usepackage{psfig}

\makeatletter
\@addtoreset{equation}{section}

\begin{document}

\newcommand{\be}{\begin{equation}}
\newcommand{\ee}{\end{equation}}
\newcommand{\tr}{{\rm tr}}
\newcommand{\ym}{{{}_{Y\! M}}}
\newcommand{\Z}{{\bf Z}}
\newlength{\extraspace}
\setlength{\extraspace}{1.5mm}
\newlength{\extraspaces}
\setlength{\extraspaces}{1.5mm}
\addtolength{\abovedisplayskip}{\extraspaces}
\addtolength{\belowdisplayskip}{\extraspaces}
\addtolength{\abovedisplayshortskip}{\extraspace}
\addtolength{\belowdisplayshortskip}{\extraspace}
\begin{titlepage}
\begin{center}

\hfill PUPT-2001\\

\bigskip
\bigskip

\vspace{6\baselineskip}

{\Large Duality Cascade and Oblique Phases in\\[3.5mm]
Non-Commutative Open String Theory}\\

\bigskip
\bigskip
\bigskip

{\large Chang S. Chan{}$^a$, Akikazu Hashimoto${}^b$, and Herman Verlinde${}^a$\\[5mm]}

{\it ${}^a$ Physics Department, Princeton University, Princeton, NJ 08544}\\[2mm]
{\it ${}^b$ Institute for Advanced Study, Princeton, NJ 08540}\\[1.8cm]

\bigskip

{\bf Abstract}

\end{center}

\noindent We investigate the complete phase diagram of the decoupled
world-sheet theory of $(P,Q)$ strings. These theories include 1+1
dimensional super Yang-Mills theory and non-commutative open string
theory.  We find that the system exhibits a rich fractal phase
structure, including a cascade of alternating supergravity, gauge
theory, and matrix string theory phases.  The cascade proceeds via a
series of $SL(2,\Z)$ S-duality transformations, and depends
sensitively on $P$ and $Q$. In particular, we find that the system may
undergo multiple Hagedorn-type transitions as the temperature is
varied.

\vfill
\noindent July 2001
\end{titlepage}
\newpage

\section{Introduction}

Duality is a powerful tool for analyzing dynamical aspects of field
theories and string theories. Some dualities are exact. They assert
that two seemingly different theories are physically equivalent.
String theory with sufficiently many unbroken supersymmetries are
believed to exhibit such exact duality relations, in the form of
discrete symmetry relations acting on the space of couplings. These
symmetries, and the fact that a large class of field theories can be
formulated as decoupling limits of string theories, have been used to
derive many examples of field theory dualities, as well as dual
correspondences between field theories and supergravity theories.
Similar considerations led to the understanding of duality relations
among less familiar theories such as non-commutative gauge theories,
non-commutative open string theories, and little string theories.

In this article, we consider the duality relations of non-commutative
open string theories \cite{ncos1,ncos2,Barbon:2000sg} in 1+1
dimensions with 16 supercharges. These theories can be formulated as a
decoupling limit of bound states of D-strings and fundamental strings
in type IIB string theory. The relevant duality relations follow from
the $SL(2,\Z)$ S-duality symmetry of type IIB string theory and the
gauge theory/supergravity correspondence applied to $(P,Q)$
strings. Our aim is to use a combination of both these dualities to
gain insight into the thermodynamic properties of the theory.  Various
authors have also considered aspects of S-duality in the context of
NCOS theories
\cite{Russo:2000zb,Russo:2000mg,Cai:2000yk,Lu:2000vv,Gran:2001tk} and
their supergravity duals \cite{Sahakian:2000ay,Harmark:2000wv}.

NCOS theory is closely related to ordinary 1+1-d super Yang-Mills
theory in at least two different ways. First, for rational value of
its string coupling $G_o^2$, it's known to be S-dual to ordinary SYM
theory with a non-zero electric flux, which therefore provides the
proper ultraviolet definition of the theory. On the other hand, like
any other known open string theory, NCOS theory reduces, at scales
sufficiently below its string scale, to an effective low energy gauge
theory. Near the NCOS string scale, however, this gauge theory
description breaks down, and the system undergoes a phase transition
into an effective matrix string theory phase \cite{ggkrw,
Barbon:2001tm}. The formation of the matrix strings can be viewed as
an ionization process of the non-commutative open strings, that escape
via the Coulomb branch from the bound state with the D-strings.

On general grounds, different dual descriptions are never
simultaneously weakly coupled, since two distinct weakly coupled
theories are manifestly inequivalent.  This means that for given
temperature and couplings, one can expect that, among the set of
theories related by duality, there typically exists one preferred
description which is most weakly coupled. In this paper, we will use
this intuition to map out the complete phase diagram of the
1+1-dimensional NCOS theory.  The main new conclusions of our study
are the following:

\begin{itemize}

\item{ In most studies done so far of the thermodynamics of
1+1-dimensional NCOS theory, the effective open string coupling
constant was assumed to be fixed at some rational value $G_o^2 =
P/Q$. We will find, however, that $G_o^2$ can be any real positive
number. This allows us to view NCOS theory as a continuous
non-commutative deformation of ordinary gauge theory.}

\item{ Upon systematic consideration of the role of full $SL(2,\Z)$
duality structure and the AdS/CFT correspondence, a remarkably
elaborate phase structure emerges. Various $SL(2,\Z)$ dual
descriptions become preferred in disjoint regions of the phase diagram
parameterized by temperature $T$ and the NCOS coupling constant
$G_o^2$.  These regions form a complicated fractal pattern.}

\item{ As a function of $T$, the theory can go through a cascade of
alternating supergravity, gauge theory, and matrix string theory
phases.  The cascade proceeds via a series of $SL(2,\Z)$ S-duality
transformations, and depends sensitively on $P$ and $Q$. In
particular, we find that the system may undergo a sequence of
successive ionization and recombination transitions. }

\end{itemize}

The fractal pattern seen in the phase diagram closely resembles the
phase structure found for the supergravity duals of non-commutative
Yang-Mills theories on a torus \cite{akisunny1,epr}.  There, the role
of $SL(2,\Z)$ was played by the Morita/T-duality group. The duality
cascade, which involved only the supergravity descriptions, were found
not to give rise to any observable thermodynamic effects, simply
because the area of the horizon in Einstein frame is invariant under
T-duality transformations.  Here, the duality cascade will act among
the gauge theory, matrix theory, and supergravity phases, giving rise
to thermodynamically observable cross-over effects.

The organization of this paper is as follows. In section 2, we collect
various preliminary facts regarding NCOS theory as decoupling limit of
string theory and the form of the $SL(2,\Z)$ S-duality
transformations.  In section 3, we analyze the role of supergravity
dual and the Hagedorn transition for the theory corresponding to a
given set of charges $P$ and $Q$. In section 4, we describe how the
various dual descriptions fit together to form a continuous, though
fractal, phase diagram. We conclude in section 5.

\section{Preliminaries}

\subsection{Parameters of 1+1-d NCOS theory}

In this subsection we introduce the parameter space of 1+1-d
non-commutative open string theory.

Since 1+1-d NCOS theory is an interacting theory of open strings, it
is specified by an open string coupling constant, $G_o$, and by the
string tension, $\alpha'_{ef\! f}$. In addition, we can introduce
$U(Q)$ Chan-Paton factors, as well as turn on a discrete electric
flux. In two dimensions, this flux behaves like a discrete $\theta$
parameter \cite{Witten:1979ka}
\begin{equation}
\theta = {2 \pi P \over Q}
\end{equation}
where $P$ is an integer ranging from zero to $Q-1$.  In the language
of the underlying IIB string theory, $P$ and $Q$ count the number of
fundamental and Dirichlet strings, respectively, that make up the
bound state \cite{Witten:1996im}.

Just as in ordinary open string theory, we can expect that, in a
suitable low energy regime, NCOS theory reduces to 1+1-d super
Yang-Mills theory with $U(Q)$ gauge symmetry.  The dimensionful gauge
coupling $g_\ym$ is related to the NCOS coupling and string length
via\footnote{Here and hereafter, we will ignore constant numerical
factors of order one.}
\begin{equation}
g_\ym^2 = {G_o^2  \over \alpha'_{ef\! f}} \ . \label{gym}
\end{equation}
This infrared gauge theory should not be confused with the S-dual
Yang-Mills theory in the ultraviolet of NCOS frequently discussed in
the literature \cite{Verlinde:1997fx,Klebanov:2000pp,
Gopakumar:2000ep}.

Since $g_\ym^2$, $\alpha'_{ef\! f}$, and $G_o^2$ are related by
(\ref{gym}), any two out of the three can be taken as the parameters
defining the theory. For our purposes it will be convenient to think
of NCOS theory as a modification of super Yang-Mills theory in the
ultraviolet, induced by an irrelevant space-time non-commutativity
perturbation.  To emphasize this point of view, we will choose our
parameters to be $g_\ym^2$ and $G_o^2$. We will see later that,
contrary to some claims in the literature
\cite{Klebanov:2000pp,Gopakumar:2000ep}, $G_o^2$ can in fact take on
arbitrary positive real values, which from the gauge theory
perspective sets the scale of the non-commutativity parameter, in
units set by $g_\ym^2$. Ordinary super Yang-Mills theory is recovered
in the limit $G_o^2 \to\! 0$, with $g_\ym$ held fixed.

To summarize, the set of independent parameters that we will use to
parameterize non-commutative open strings in 1+1 dimensions are the
following:
\begin{equation}
\left\{\rule{0ex}{2ex} g_\ym^2, \quad G_o^2, \quad P, \quad Q\right\}\ .
\end{equation}
For the purpose of studying the action of $SL(2,\Z)$ duality group on
these parameters, it will sometimes be useful to separate the factor
$N$ which is the greatest common divisor of $P$ and $Q$ and write
\begin{equation}
P = Np, \qquad Q = Nq
\end{equation}
where $p$ and $q$ are relatively prime integers.

\subsection{SYM Decoupling limit of (P,Q) strings}

Here we recall the decoupling limit of the $(P,Q)$ string theory that
produces 1+1-dimensional SYM theory, and introduce its supergravity
dual.

Starting from the world sheet theory of a $(P,Q)$ string bound state
in IIB string theory, we can consider the limit
\begin{equation}
g_s \rightarrow 0 \label{scaling}
\end{equation}
while focusing on physics taking place at energy scale or temperature
of order
\begin{equation}
T^2 \sim g_\ym^2 = {g_s \over \alpha'} \ . \label{focus}
\end{equation}
In this limit, the world sheet theory reduces to 1+1 dimensional super
Yang-Mills theory with gauge coupling $g_\ym$, gauge group $U(Q)$ and
$P$ units of electric flux.

To formulate the corresponding near horizon supergravity geometry, we
can start from the full IIB supergravity solution of the $(P,Q)$
string obtained by Schwarz in \cite{Schwarz:1995dk}. This solution is
parameterized by the asymptotic values of the string coupling and
axion field, and by the two quantized charges $P$ and $Q$.  Applying
the scaling limit (\ref{scaling}), while focusing on the range of
radial coordinates parameterized by $U = {r / \alpha'}$ with $r$ as in
\cite{Schwarz:1995dk}, gives
\begin{eqnarray} 
ds^2 & = & \alpha' \left( { U^3 \over g_\ym \sqrt{Q}} (-dt^2 + dx^2) +
{g_\ym \sqrt{Q} \over U^3} (dU^2 + U^2 d \Omega_7^2) \right)
\label{sg1}\\[2mm]
e^\phi & = &  {g_\ym^3 \sqrt{Q} \over U^3} \\[2mm]
\chi & = & {P \over Q} \label{chi1}\\[2mm]
B_{NS} &=& 0 \\[2mm]
B_{RR} &=& {\alpha' U^6 \over g_\ym^4 Q} \ .\label{sg2}
\end{eqnarray}
The metric, the dilaton, and the two-form fields are exactly the same
as the near horizon limit of the $(0,Q)$ string \cite{imsy}.  The
effect of the non-vanishing electric flux manifests itself only in the
constant axion background (\ref{chi1}).

The dual supergravity description of the $(P,Q)$ gauge theory is valid
in the regime of couplings and scales where both the string coupling
$e^\phi$ and the curvature of the near-horizon geometry, measured in
string units, remain small.

\subsection{The NCOS decoupling limit of (P,Q) strings}

Here we describe the decoupling limit of $(P,Q)$ string theory that
produces 1+1-dimensional NCOS theory, and introduce its supergravity
dual. A new element in our discussion is that, as a result of
including the IIB axion field, the NCOS coupling $G_o^2$ is
incorporated as a continuous free parameter.

1+1-dimensional NCOS theory arises from the world-sheet theory on the
$(P,Q)$ string bound state in IIB theory upon taking the decoupling
limit
\be
\label{decncos}
g_s \to \infty,
\qquad \quad  \, {g_s^2\, \alpha'  } \ \ {\rm fixed}.
\ee
This limit is S-dual to the SYM limit (\ref{scaling}).  The NCOS
parameters $\alpha'_{ef\! f}$ and $G_o^2$ are related to the IIB
parameters via
\be
\label{relations}
g_s^2\, \alpha' = G_o^4 \,  \alpha'_{ef\! f}, \,
\qquad \quad\alpha' \tr F = 1- {\alpha' \over 2 \alpha'_{ef\! f}}, \ee
where $F= \epsilon^{01} {F_{01}}$ is the $U(Q)$ Yang-Mills field
strength on the D-string worldsheet.  In the limit (\ref{decncos}),
$\tr F$ is automatically tuned to approach its critical value $\alpha'
\tr F = 1$, at precisely such a rate that its electrostatic force
counteracts the infinite fundamental open string tension, so as to
produce a finite effective tension $\alpha'_{ef\! f}$ of the NCOS
strings.

To see this explicitly, consider the Born-Infeld effective lagrangian
of the D-string bound state (omitting all fields except the 1+1-d
gauge field)
\be
{\cal L} = -{1\over  g_s \alpha'  } \tr \sqrt{( 1 - ( \alpha' F)^2 )} + 
\chi \tr F.
\ee
Here we included the topological term associated with the constant
axion field $\chi$.  The compactness of the gauge group implies that
the $U(1)$ part of the electric field $P = \tr E$ where $E$, defined
as the canonical conjugate to the gauge field $E= {\partial {\cal L}
\over \partial \dot{A}}$, takes on integer values only. Inverting the
relation
\be P - \chi Q = { \alpha' \tr F \over g_s \sqrt{(1 - ( \alpha'
F)^2)}} \ee
reveals that the field strength indeed becomes near-critical in the
NCOS limit (\ref{decncos})
\be  \tr F \, \simeq\, 1 - {Q^2 \over 2 g_s^2 (P\! -\! \chi Q)^2}.
\ee
Furthermore, from the relations (\ref{relations}) defining the NCOS
parameters, we read off that
\be
\label{gnot}
G_o^2 = {Q \over |P\! -\! \chi Q|}.
\ee
As claimed, the effective coupling $G_o^2$ can thus indeed attain
arbitrary real, positive values.  As we will see shortly, the
existence of more general NCOS theories with a continuously varying
coupling also naturally arises from S-duality symmetry of the
underlying IIB string theory.

The full supergravity solution dual to the 1+1-d non-commutative open
string theory, for arbitrary values of the parameters $\Bigl\{ g_\ym,
\, G_o,\, P,\, Q\, \Bigr\}$, follows from the general expression
obtained by Schwarz in \cite{Schwarz:1995dk}, by applying the NCOS
scaling limit (\ref{decncos}). One finds
\begin{eqnarray}
ds^2 &=& \alpha' \left(1 + {U^6 G_o^4 \over g_\ym^6 Q }\right)^{1/2}
\left( {U^3 \over g_\ym\sqrt{Q} } (-dx_0^2 + dx_1^2) + {g_\ym
\sqrt{Q}\over U^3} (dU^2 + U^2 d \Omega_7^2)\right), \label{sg0}
\\[2mm]
e^\phi &=& {g_\ym^3 \sqrt{Q} \over U^3} \left(1 + {U^6 G_o^4 \over
g_\ym^6 Q }\right), \\[2mm]
\chi &=& {g_\ym^2 P Q + G_o^2 (G_o^2 P + Q) U^6 \over Q (g_\ym^6 Q +
G_o^4 U^6)}, \\[2mm]
B_{NS}&=& - {\alpha' G_o^2U^6 \over g_\ym^4 Q} , \\[2mm]
B_{RR} & =& {\alpha' (Q + G_o^2 P) U^6 \over g_\ym^4 Q^2} \  .
\end{eqnarray}
Here, relative to the notation used in \cite{Schwarz:1995dk}, we made
the identifications
\begin{equation}
g_s^2 = {G_o^6 \over \alpha' g_\ym^2} , \qquad 
\chi_\infty = {P \over Q} + {1 \over 
G_o^2}, \qquad r^2 = {\alpha' G_o^2 U^2 \over g_\ym^2} \label{scaling2}.
\end{equation}
This dual supergravity description of $(P,Q)$ NCOS theory is valid in
the regime of couplings and scales where both $e^\phi$ and the
curvature of the near-horizon geometry remain small.

\subsection{SL(2,\Z) duality of NCOS theory}

In this subsection, we describe how $SL(2,\Z)$ S-duality
transformations act on the NCOS data $g_\ym^2$, $G_o^2$, $P$, and $Q$,
and write the supergravity data in a more manifestly S-duality
covariant form.

Before taking any decoupling limit, the $(P,Q)$ strings are permuted
by the $SL(2,\bf{Z})$ S-duality symmetry of the IIB theory, via
\begin{equation}
\label{sone}
\left( \begin{array}{c} \tilde{P} \\ \tilde{Q}      \end{array}\right) =
\left(\begin{array}{cc} a & b \\ c & d \end{array}\right)
\left(\begin{array}{c} P \\ Q \end{array}\right),
\end{equation}
which leaves $N\!=\,$gcd$(P,Q)$ invariant. The string coupling $g_s$
and axion $\chi$ transform via
\be
\label{lambda}
\tilde{\lambda} = {a {\lambda} + b \over c {\lambda} + d} \, ,
\qquad \quad \lambda = \chi + {i\over g_s}.
\ee
Evidently, S-duality does not preserve the super Yang-Mills decoupling
limit; instead it is mapped onto the NCOS limit.

The NCOS limit, on the other hand, is in general preserved.  The axion
field $\chi_\infty$ transforms like
\begin{equation}
\label{chitra}
\tilde{\chi}_\infty  = {a \chi_\infty + b \over c \chi_\infty + d}
\end{equation}
in the scaling limit. For generic values of $\chi_\infty$, the
$SL(2,\Z)$ transformed value is again finite, and thus the transformed
theory is also a regular NCOS theory. Using the relations
(\ref{relations}) and (\ref{gnot}) with the IIB parameters, a
straightforward calculation shows that the $SL(2,\Z)$ transformation
law for the NCOS parameters reads
\begin{eqnarray}
\left(\begin{array}{c}\tilde P \\ \tilde Q \end{array}\right) &= &
\left(\begin{array}{cc}a & b \\ c & d \end{array}\right) 
\left(\begin{array}{c} P \\  Q \end{array}\right) \\
\tilde g_\ym^2 & = & {g_\ym^2 (c P + d Q)^3 \over Q^3}, \label{trans3} \\
\tilde G_o^2 & = & {(c P + d Q) (c Q + (c P + d Q) G_o^2) \over Q^2}. \label{trans4}
\end{eqnarray}
These formulas closely resemble the Morita duality transformations of
the parameters of non-commutative Yang-Mills theory
\cite{akisunny1,Pioline:1999xg}.

Note that, in the special case that $\chi$ and $G_o^2$ are rational,
there is always one particular $SL(2,Z)$ transformation for which the
denominator in (\ref{chitra}) vanishes, which implies that the
transformed theory has $\tilde{G}_o^2\!=\! 0$. Hence in this case the
NCOS theory can be mapped back onto a commutative SYM theory.
Conversely, this means that any NCOS theory with rational $G_o^2$ has
a precise field theoretic definition via this equivalent SYM gauge
theory. NCOS theories with irrational $G_o^2$, on the other hand, do
not have such a UV definition; they need to be defined via the
corresponding IIB decoupling limit (\ref{decncos}).

In order to make the $SL(2,\Z)$ multiplet structure of the
supergravity background more manifest, it is convenient to go to the
Einstein frame, where the metric becomes
\be
ds^2 =  l_p^2 \left({U^3 \over g_\ym^3 \sqrt{Q}}\right)^{1/2}
\left( {U^3 \over g_\ym\sqrt{Q} } (-f(U)dx_0^2 + dx_1^2) 
+ {g_\ym
\sqrt{Q}\over U^3} (f^{-1}(U) dU^2 + U^2 d \Omega_7^2)\right).
\ee
Here we have included the thermal factor
\begin{equation}
f(U) = 1 - {U_0^6 \over U^6} \ .
\end{equation}
This solution describes a non-extremal black string with a horizon
located at $U = U_0$. It should therefore be interpreted as the
supergravity dual of generic NCOS theory at a finite temperature
$T_0$.  The relation between $U_0$ and $T_0$ can be determined by
analytically continuing the solution to Euclidean signature $t = i
\tau$ and looking at the metric in the $U$ and $\tau$ coordinates near
$U = U_0$. Introducing the coordinate
\begin{equation}
\rho^2 \sim 1 - {U_0^6 \over U^6}
\end{equation}
the metric near $\rho=0$ take the form
\begin{equation}
ds^2 \sim (d\rho^2 + {9 U_0^4 \over g_\ym^2 Q} \rho^2 d\tau^2)
\end{equation}
from which we infer that (up to factors of order one), the temperature
$T_0$ equals
\begin{equation}
T_0 \simeq {U_0^2 \over g_\ym \sqrt{Q}} \ .
\end{equation}
This is the standard  UV/IR relation for D1-branes \cite{Peet:1998wn}.

Since temperature is a physical notion independent of the S-duality
orbit, let us now choose to parameterize the radial coordinate by
\begin{equation}
T =  {U^2 \over g_\ym \sqrt{Q}} \ ,
\end{equation}
and introduce the $SL(2,\Z)$ invariant combination
\begin{equation}
\gamma^2 = {g_\ym^2 \over Q^3}.
\end{equation}
Replacing $g_\ym^2$ and $U$ by $\gamma^2$ and $T$, the supergravity
solution in Einstein frame becomes
\begin{eqnarray} 
ds^2 & = & l_p^2 \left( {T \over \gamma} \right)^{1/4} \left( T^2
(-dt^2+dx^2) + {1 \over 4T^2} dT^2 + d \Omega_7^2\right)
\label{metric} \\[2mm]
\left(\begin{array}{c}B_{NS} \\ B_{RR} \end{array}\right) & = & l_p^2
\left(\begin{array}{c}-G_o^2 \\ 1+ {G_o^2 P \over Q}
\end{array}\right) \left({T^3 \over \gamma Q}\right) \\[2mm]
e^\phi & = &{ \gamma^3 Q^4 + G_o^4 T^3 \over \gamma^{3/2} Q^2 T^{3/2}}
\label{dilaton}\\[2mm]
\chi & = & {\gamma^3 P Q^4 + G_o^2(G_o^2 P + Q) T^3 \over Q( \gamma^3
Q^4 + G_o^4 T^3)} \ .
\end{eqnarray}
As expected, we see that the Einstein frame metric is S-duality
invariant.  The two-form fluxes and the axion and dilaton fields, on
the other hand, transform covariantly.

In the following, we will identify the dilaton profile (\ref{dilaton})
as a function of the radial coordinate $T$ with the actual effective
string coupling, in the given $(P,Q)$ frame, as a function of the
physical temperature. The justification for this identification is
that, at temperature $T_0$, most of the degrees of freedom of the
supergravity can be thought of as being localized near the black
string horizon at $T=T_0$. Indeed, provided we are in a regime where
the supergravity description is valid, we can identify the
thermodynamic entropy density of the NCOS theory with the
Bekenstein-Hawking entropy of the black string
\begin{equation}
s = {S \over V} = {A_H \over l_p^8 V } = {T^2 \over \gamma}. \label{entropy}
\end{equation}
This confirms that the NCOS matter can be thought of as forming the
thermal atmosphere of the black string, and that the strength of
interactions is governed by the effective string coupling $e^\phi$
close to the horizon at $T=T_0$.

\section{Phases of NCOS theory}

In the previous section, we formulated NCOS both as a decoupled theory
on a brane and as a supergravity dual.  As we emphasized in the
introduction, these two formulations of the same theory should
complement one another, in the sense that depending on the
circumstances, one or the other should single itself out as the
preferred description of the system. Let us address this issue
concretely by first fixing all of the parameters $g_\ym^2$, $G_o^2$,
$P$ and $Q$, while varying the temperature.

For starters, the value of the dilaton needs to stay small in order
for the theory to be weakly coupled. From the form of the dilaton
given in (\ref{dilaton}), we find that this restricts the temperature
to take value in the range $e^\phi \ll 1$, or
\begin{equation}
 Q^2 \left({1 - \sqrt{1 - 4 G_o^4}  \over 2 G_o^4} \right)
\ll \left({T\over \gamma}\right)^{3/2} \ll  Q^2  \left({1 + \sqrt{1 - 4
G_o^4} \over 2 G_o^4}\right) \ . \label{region1}
\end{equation}
It is clear that $G_o^2$ must be less than $1/2$ for this region to
exist, and that the region is bounded below at $T^{3/2} = \gamma^{3/2}
Q^2$. We will analyze what happens outside this range of temperatures
in the following section.

\begin{figure}
\centerline{\psfig{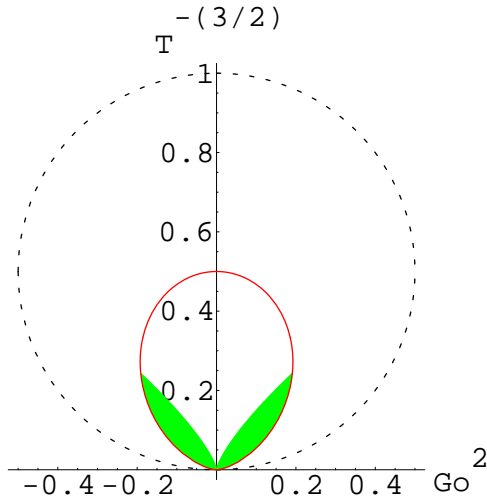}}
\caption{\label{figa} This figure indicates the regimes of validity of
the three possible phases of NCOS theory, for given charges $P$ and
$Q$: (i) the supergravity phase, inside the black dashed circle,
outside the red line, (ii) the gauge theory phase, inside the red
line, and (iii) the matrix string phase, the green shaded region.}
\end{figure}

Let us now explore the full range of validity of the $(P,Q)$
supergravity description. The general criteria for the effectiveness
of supergravity description \cite{imsy} dictate that the curvature
radius as measured in the string frame metric should be large compared
to the string length. The curvature radius of the background
(\ref{metric}) can be estimated from the radius of the 7-sphere
forming the black string horizon.  Comparing this radius with the
string length, we thus deduce that the supergravity approximation
breaks down in the region
\begin{equation}
{R^2 \over \alpha'}  = \left({T \over \gamma}\right)^{1/4} e^{-\phi/2}  \ll 1\ . \label{region2}
\end{equation}
It can be seen that this region is completely contained inside of the
range (\ref{region1}). Both regions are indicated in figure
\ref{figa}, for values of $G_o^2$ ranging from $-1/2$ to $1/2$. For
the vertical coordinate in the figure we have chosen $T^{-3/2}$, so
that the $e^\phi = 1$ boundary (\ref{region1}) takes the form of a
circle, indicated by the black dashed line. The regime (\ref{region2})
is bounded by the red solid line, so that the supergravity description
is valid in the region inside the black dotted line and outside the
red solid line.  Note that the ultraviolet is at the lower end of the
figure.

Since the curvature radius of the supergravity is small inside the red
circle, we can expect that the dual gauge theory description may take
over in this region. This is most easily verified at the special
vertical line at $G_o^2 = 0$, where the non-commutativity parameter is
turned off.  This line corresponds to ordinary 1+1 dimensional super
Yang-Mills theory.  The red line intersects the $G_0^2 = 0$ axis at
$T^2 = g^2_\ym/Q$, which is indeed exactly the point where the 't
Hooft coupling of the gauge theory is of order one. Moreover, since
SYM theory in 1+1 dimensions is super-renormalizable, the gauge theory
description remains valid for arbitrarily high energies; this is
indicated in the figure \ref{figa} by the fact that the red circle is
touching the abscissa.

Away from $G_o^2=0$, the effect of non-commutativity should manifest
itself. Specifically, when starting from the infra-red (from above in
the figure), we expect that when the temperature $T$ reaches the scale
set by the NCOS string tension
\begin{equation} T = {1 \over
\sqrt{\alpha'_{eff}}} = \sqrt{{g_\ym^2 \over G_o^2}} \end{equation}
the theory must undergo a Hagedorn transition. Beyond this
temperature, the system is most accurately described by a matrix
string theory (MST) phase \cite{ggkrw}, that is a sigma model on ${\bf
R}^8/S_N$ describing the eigenvalue dynamics of the matrix scalar
fields of the SYM model.  These matrix strings can be thought of as
ionized NCOS strings, that due to the thermal fluctuations have
managed to escape the D-string bound state via the Coulomb branch.  A
concrete quantitative check of this physical picture is provided by
the fact that the effective tension of a long fundamental string that
escapes to infinity in the supergravity geometry (\ref{sg0}) coincides
with the tension of the NCOS strings:
\be {1 \over 2 \alpha'_{ef\! f}} = {1 \over \alpha'} (\sqrt{g^s_{00}
g^s_{11}} + B_{NS}) = {g_\ym^2 \over 2 G_o^2}. 
\ee
The range of temperatures in which this phase dominates is illustrated 
by the green shaded region in figure \ref{figa}.

At even higher temperatures, one hits the boundary of the red region
where the supergravity description again becomes valid. At that
temperature, the long NCOS strings recombine with the D-string due to
the strong gravitational attraction caused by the black hole geometry
of the supergravity dual \cite{Hashimoto:2000ys}. The sequence of
phases
\begin{equation}
\mbox{SUGRA} \rule{4ex}{0ex}  \rightarrow \rule{4ex}{0ex} 
\mbox{NCOS}  \rule{4ex}{0ex}  \rightarrow \rule{4ex}{0ex} 
\mbox{MST}   \rule{4ex}{0ex}  \rightarrow \rule{4ex}{0ex}   
\mbox{SUGRA} \label{chain}
\end{equation}
going up in temperature was also described in \cite{ggkrw}. 

\section{SL(2,\Z) duality cascades}

Our remaining task now is to describe what happens outside the black
circle in figure \ref{figa}. Clearly, since the effective string
coupling is getting large there, we can expect that the system goes
over into another S-dual regime. A small subtlety is that, because of
the non-trivial axion background, a simple inversion will not
necessarily map strong coupling to weak coupling. More general
$SL(2,\Z)$ transformations may be needed.

To address this issue in a systematic way, we will take advantage of
the survey of $SL(2,\Z)$ duality transformations of NCOS theory given
in section 2. To begin, it will turn out to be convenient to exploit
the S-duality equivalence and combine all theories into one single
parameterization. The most convenient choice is to take as the base
theory, the system with charges
\begin{equation}
P=0, \qquad Q=N
\end{equation}
and couplings
\begin{equation}
g_\ym^2, \qquad G_o^2,
\end{equation}
and parameterize all the dual theories by the element
\begin{equation}
\Lambda = \left(\begin{array}{cc} a & b \\ c & d \end{array}\right) \in  SL(2,\Z)
\end{equation}
which maps it to the system with couplings and charges
\begin{equation}
\tilde g_\ym^2 = d^3 g_\ym^2, \qquad \tilde G_o^2 = d (c+d G_o^2),
\qquad \tilde P = b N, \qquad \tilde Q = d N \ .
\end{equation} 
In other words, we can use $g_\ym^2$ to set the scale, and $G_o^2$ as
the data parameterizing the $SL(2,\Z)$ equivalence class, and $c$ and
$d$ as the data parameterizing the specific elements of the $SL(2,\Z)$
orbit.

In terms of these data, the dilaton profile (\ref{dilaton}) as a
function of temperature takes the following form
\begin{equation}
e^\phi = (c + d G_o^2)^2 \left({T^{3/2} \over \gamma^{3/2} N^2}\right)
+ d^2 \left({\gamma^{3/2} N^2 \over T^{3/2}} \right) \ .
\end{equation}
At this point, it is convenient to introduce the dimensionless
parameters
\begin{equation}
x = G_o^2, \qquad y = {\gamma^{3/2} N^2 \over T^{3/2}}  \label{dilaton2}
\end{equation}
to quantify the coupling and the temperature, respectively. Note, as
we did in section 4, that $y$ scales like $T^{-3/2}$ so small $y$
corresponds to large temperature. In terms of $x$ and $y$, the dilaton
profile (\ref{dilaton2}) becomes
\begin{equation}
e^{\phi} = {(c + d x)^2 \over y} + d^2 y \ . \label{dilaton3}
\end{equation}

\begin{figure}
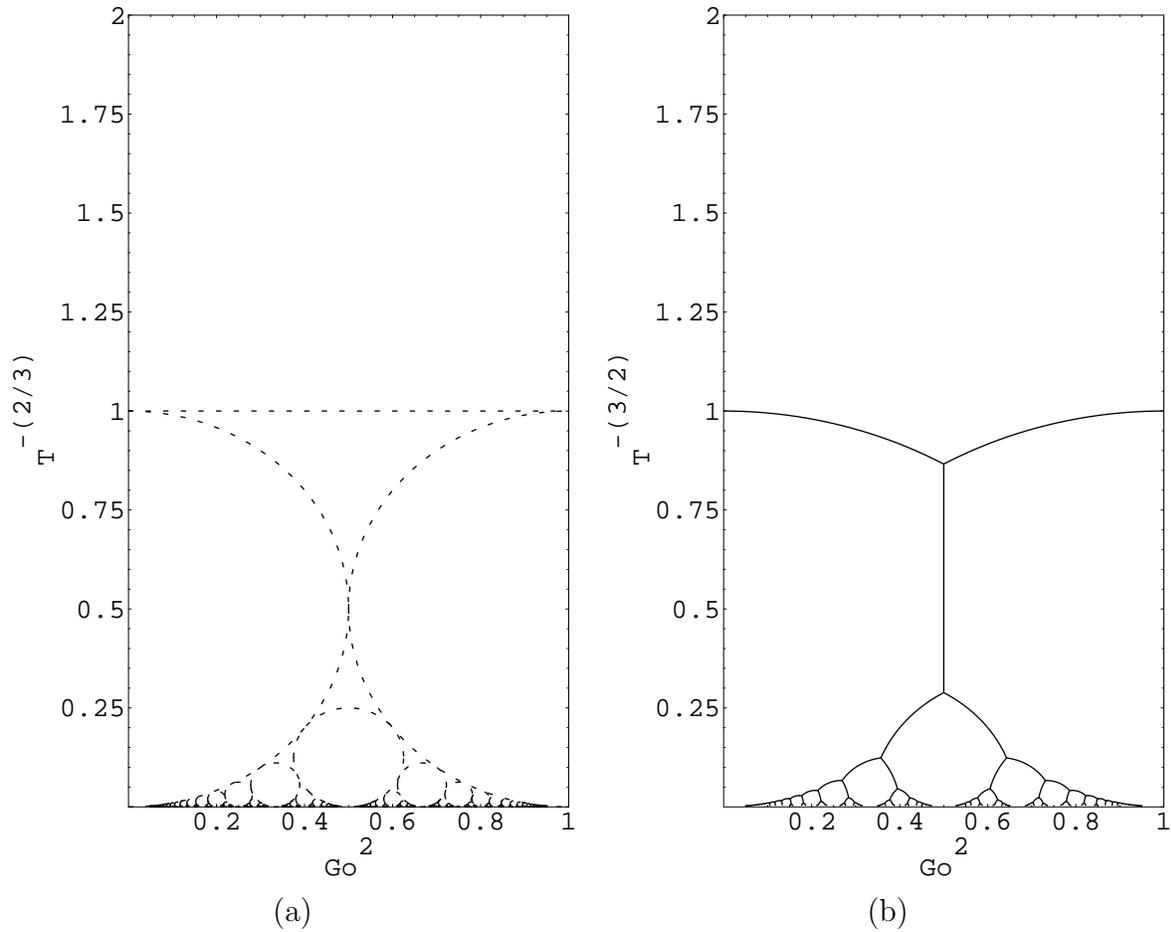

\begin{tabular}{cc}
\psfig{file=ncos1d4.epsi,width=2.95in}  &  \psfig{file=cardy2.epsi,width=2.95in} \\
(a) & (b)
\end{tabular}
\caption{\label{figb} In the left figure, we have indicated the
circles in the $(x,y)$ plane inside of which $e^\phi<1$ for some
integers $(c,d)$.  In the right figure, these regions are extended,
such that the adjoining $(c,d)$ cells have the same value for the
dilaton along the boundary.  Inside each cell, one unique $(c,d)$
description minimizes the dilaton.}
\end{figure}

Our task is to determine, for given value of the parameters $x$ and
$y$, which effective theory provides the best description of the
system.  As a first step in this procedure, we will identify the pair
of integers $(c,d)$ which minimizes the string coupling
(\ref{dilaton3}) at each given point in the $(x,y)$-plane. For this
purpose, it is helpful to first draw the locus on the $(x,y)$-plane
for which $e^\phi = 1$ for all possible integers $(c,d)$. These loci
are circles and are illustrated in figure \ref{figb}.a.

The circle corresponding to $(c,d)= (0,1)$ is the one drawn earlier in
figure \ref{figa}; the rest are its generalizations to other values of
$(c,d)$. Inside each of the circles, the corresponding string coupling
$g_s = e^\phi$ is smaller than 1. None of the circles overlap, so if a
point $(x,y)$ happens to be inside a $(c,d)$ circle, the most weakly
coupled theory is the one labeled by $(c,d)$.  There are some points,
however, which are not covered by the circles. Here we can not apply
$SL(2,\Z)$ duality to make the string coupling less than one. However,
since we are only interested in identifying the dual theory which
minimizes the dilaton, we can take the freedom to extend the circular
regions in such a way that adjoining $(c,d)$ cells have the same value
of the dilaton along the boundary. The resulting $(c,d)$ cells, which
now fill the entire $(x,y)$ plane, are illustrated in figure
\ref{figb}.b.

These $(c,d)$ regions on the phase diagram have an identical structure
to that found in \cite{epr} in the context of non-commutative gauge
theory on a torus, where the role of $SL(2,\Z)$ was played by the
Morita equivalence relation.  As was emphasized in \cite{epr}, these
phase structures also bear very interesting resemblance to the phase
structure of lattice spin models with $\theta$ parameters considered
in \cite{Cardy:1982qy,Cardy:1982fd}. Similar structures have also
appeared in the context of dissipative Hofstadter model
\cite{Callan:1992da} and quantum Hall systems \cite{Fradkin:1996xb}.

\begin{figure}
\centerline{\psfig{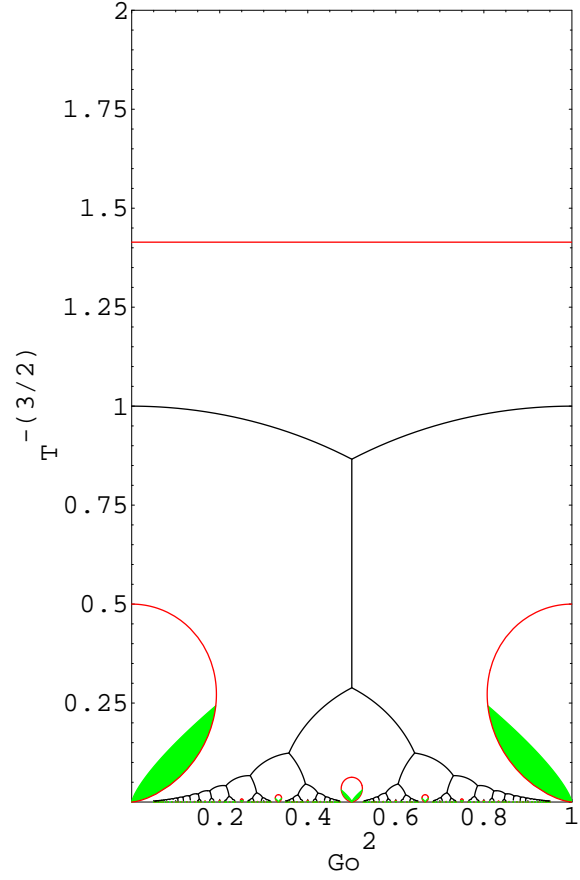}}
\caption{\label{figc} The phase diagram that combines all possible
phases of SYM/NCOS theory in 1+1 dimensions for fixed $N=\gcd(P,Q)$. A
unique $({P},{Q})$ theory provides the most weakly coupled description
inside each fundamental domain. Each fundamental domain is further
divided into the supergravity phase (outside the red circle), the
field theory phase (inside the red circle), and the coexisting matrix
string phase (shaded green region).}
\end{figure}

The analysis of the phase structure in each of the $(c,d)$ cells will
closely parallel our earlier discussion in section 3. In particular,
we expect that within each of the cells, we can identify three
different regions, corresponding to the effective gauge theory phase,
matrix string phase, and the supergravity phase. The respective ranges
of validity of these different phases are summarized in the phase
diagram displayed in figure \ref{figc}.

Let us highlight some of the features of this phase diagram.
\begin{itemize}
\item The vertical axis is proportional to $T^{-3/2}$, so that the
ultraviolet corresponds to the bottom, and the infrared to the upper
end of the figure.  Each vertical slice corresponds to the $(0,N)$
NCOS theory with given $G_o^2$.
\item For every rational value of $G_o^2$, there is a $(c,d)$ cell
touching the horizontal axis at $y=0$. At this point
\begin{equation}
\tilde G_o^2 = d(c + d G_o^2) = 0, 
\end{equation}
it corresponds to an ordinary super Yang-Mills theory, with gauge
group $U(d N)$ and electric flux $P= bN$.

\item Starting from a given SYM gauge theory with E-flux, one can flow
upwards in $y$ toward the infrared. It is possible that the system
then crosses over into another $(c,d)$ cell, and reaches another
effective gauge theory phase.  This effective gauge theory is deformed
with an irrelevant non-commutative perturbation, proportional to the
effective tension of the corresponding NCOS phase. This sequence of
phases has been described in \cite{ggkrw}. What we see here, however,
is that the sequence of phases does not necessarily stop here: by
going further toward the infrared, one can potentially cross many more
$(c,d)$ cells before reaching the deep infrared region at $y =
\infty$.
\item The number of $SL(2,\Z)$ transformations involved in the flow
from the UV to the IR depends sensitively on the rationality of
$G_o^2$. Irrational values of $G_o^2$ require infinitely many
$SL(2,\Z)$ transformation in the ultraviolet region, as indicated by
the fractal phase pattern illustrated in figure \ref{figc}.
\item Since the ionization/recombination phase transition associated
with the Hagedorn scale takes place in each of the $(c,d)$ cells, the
system can undergo these transitions multiple times as the temperature
is varied monotonically.
\end{itemize}

As a concrete illustration of this type of duality cascade, let us
consider a given theory with parameters
\begin{equation}
P=0, \qquad Q=N, \qquad G_o^2 = {1 \over n_1 - {1 \over n_2}}
\end{equation}
corresponding in the ultra-violet to a $U(N(n_1n_2-1))$ super
Yang-Mills theory with $Nn_1$ units of electric flux.  The SYM degrees
of freedom, however, are weakly coupled in the far UV only. In
particular, since electric flux creates a mass gap in two dimensions,
one expects that towards the infra-red, the $U(Q)$ gauge symmetry gets
broken to $U(N)$.  Ultimately, the system will flow towards $U(N)$
matrix string theory.

\begin{figure}
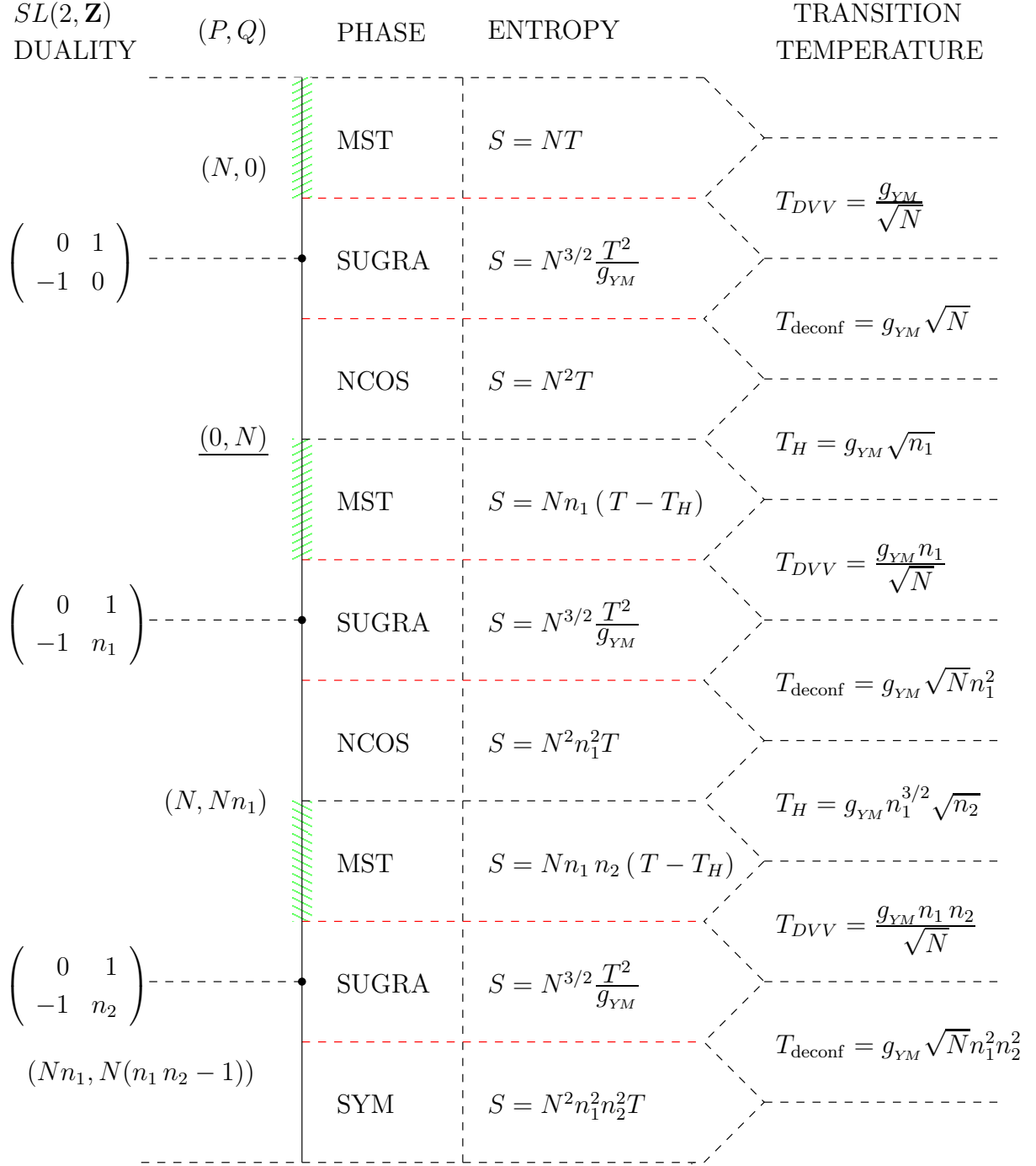

\input xbig2.pstex_t
\caption{ \label{figd} An overview of the duality cascade and all the
intermediate phases for the special case of a $(0,N)$ NCOS theory with
$G_o^2 = {1 \over n_1 - {1 \over n_2}}$ with $n_2 \gg N n_1 \gg N^2
\gg 1$. We have also given the qualitative behavior of the entropy,
and the transition temperatures. The entropy is maximal in the
ultraviolet (bottom of the figure), and decreases monotonically
towards the infrared (top of the figure).}
\end{figure}

It is instructive to trace all the intermediate phases between the UV
gauge theory phase and the IR matrix string phase. They are listed in
figure \ref{figd}, where we have also indicated the behavior of the
entropy in all different phases. It will turn out that the phases are
well separated provided that $n_2 \gg N n_1 \gg N^2 \gg 1$.

Starting from the infra-red, flowing down towards the UV, the system
first follows the successive phases outlined in \cite{ggkrw} and
\cite{imsy}. Continuing further towards the UV, however, the theory
again enters a supergravity phase. At the point where $e^\phi=1$, we
now need to apply the S-duality transformation
\begin{equation}
\left( \begin{array}{rr}0 & 1 \\ -1 & n_1 \end{array}\right),
\end{equation}
connecting to an NCOS theory with charges $(N, Nn_1)$.  Then, after a
similar sequence of supergravity, gauge theory, and matrix string
theory phases, the duality cascade continues via the S-duality
transformation
\begin{equation}
\left( \begin{array}{rr} 0 & 1 \\ -1 & n_2 \end{array}\right)
\end{equation}
which finally takes us to a commutative theory with charges $(P,Q) = (N n_1,N
(n_1 n_2 - 1))$.

\section{Conclusions}

In this article, we investigated the full phase structure of
non-commutative open string theories in 1+1 dimensions and found a
rich fractal structure closely resembling the phase structure of
non-commutative gauge theory on a torus.  The most striking conclusion
of our analysis is that, with increasing temperature, the system can
undergo multiple transitions between alternating gauge theory, matrix
string theory, and supergravity phases.

A comment is in order regarding the nature of the thermodynamic
transitions at the various phase boundaries. Since we are working in
1+1 dimensions, strictly speaking there should not be any phase
transition, unless possibly when we take the large $N$ limit.  At
large but finite $N$, it is more appropriate to refer to the
transitions between the different phases as ``crossovers.''  It is
conceivable, however, that in the $N \to \infty$ limit, some of the
crossovers (in particular the Hagedorn transition) may actually become
true phase transitions. Even without sharp transitions, however, the
duality cascade described in this article should have many observable
consequences.

For rational values of $G_o^2$, the thermodynamics described here is
that of ordinary super Yang-Mills theory with some electric flux. It
would be interesting to see if it is possible to reproduce some of our
results more directly via other techniques. For example, it may be
possible to find signatures of these phase transitions in the behavior
of the thermal partition function of the gauge theory or of the
two-point function of the stress energy tensor
\cite{Hashimoto:1999xu}.  Perhaps a computation on the lattice or DLCQ
methods \cite{Antonuccio:1999iz,Hiller:2000nf} can provide some useful
insights.

Much of the $SL(2,\Z)$ structure of NCOS theory in 1+1 dimension can
rather straightforwardly be generalized to higher dimensions. There
are several important differences, however.  In 3+1 dimensions, for
example, $g_\ym^2$ is a freely adjustable dimensionless
quantity. Therefore, the full phase diagram will be three dimensional,
parameterized by $g_\ym^2$, $G_o^2$, and $T$ (measured in units of the
NCOS string length). A preliminary study indicates that the cross
sections with constant $g_\ym^2$ display an analogous structure as
describe here, whereas the cross sections for constant $G_0^2$ look
similar to the phase diagram described in \cite{Hashimoto:2000ys} for
the case of small $G_o^2$. It should be instructive to map out the
full multi-dimensional phase structure also for these higher
dimensional cases.

Another interesting open question about the 1+1-dimensional case is
what happens in the case of irrational $G_o^2$. In this case the
system does not have any known UV definition. Nonetheless, at any
finite temperature, it can be approximated to arbitrary precision by a
sequence of rational $G_o^2$ theories, which do have a UV
description. It would be very interesting to find out whether the
irrational theory allows for an independent UV fixed point
description.

\section*{Acknowledgements}

We would like to thank 
M.~Berkooz, 
S.~Gubser, 
N.~Itzhaki, and 
E.~Rabinovici
for useful remarks and discussions.  We also thank ITP, Santa Barbara
and the Amsterdam String Workshop, where part of this work was done.
The work of AH is supported in part by the Marvin L.~Goldberger
fellowship and the DOE under grant No.\ DE-FG02-90ER40542, and that of
H.V. and C.S.C. is supported by NSF-grant 98-02484.

\bibliography{sl2}		
\bibliographystyle{utphys}

\end{document}